\documentstyle[12pt]{article}
		  
\title{ On the quantum super Virasoro algebra} 

\author{\bf {M. Mansour}\\\\ 
Laboratoire de Physique Th\'eorique, Facult\'e des Sciences
\\B. P. 1014 Rabat.  Morocco \thanks{e-mail: mansour70@mailcity.com} }

\begin{document}           
 
\maketitle

$$ABSTRACT$$
The  quantum super-algebra structure on the deformed super Virasoro algebra  
is investigated. More specifically we established the possibility of defining
a non trivial Hopf super-algebra on both one and two-parameters deformed
super Virasoro algebras.
\vfill

\pagebreak

Quantum algebras (quantum groups)\cite{r1,r2}, are generalizations of the
 usual
Lie algebras to which they reduce for appropriate values of the deformation 
parameters. From the mathematical point of view they are Hopf algebras 
\cite{r3}.

 Quantum superalgebras 
appeared naturally when the quantum inverse scattering method \cite{ks1} 
was generalized to the super-systems \cite{ks1}. Related R-matrix were
considered in \cite{kr,kt} and simple examples where
presented in \cite{ku}.   The works \cite{fs,ckl,ck} are devoted to the 
q-bosonization of the q-superalgebras. The simple example of the dual
 objects,  i.e.  quantum supergroups, where investigated in \cite{ck,pu,pp}.
  A complete classification
of finite dimensional Lie Superalgebras over $C$ has been give
by Kac \cite{k1,k2} and Scheurnert \cite{sc}.\\
Supersymmetry seems to be in the realm of almost all attemts to obtain a
 unified theory of all interactions \cite{f,v}. In  particular, 
many problems of string theory are solved once sumersymmetry is 
introduced \cite{g,s}. The interplay of supersymmetry and conformal 
invariance is an intersting application of the rules up to now.
The two-dimensional superconformal field theorie can be treated from a 
group theoretical point of view, the basic ingredient is the 
two-dimensional superconformal algebra, which is infinite dimensional 
called the super Virasoro , see \cite{rs} and references therin,  The central extension of the super Virasoro algebra leads to the so called the Ramond-Neveu-Schwarz algebras\cite{ns,r}.
A supersymmetric extension of the work of Curthright and Zachos is given in
\cite{z1} by using a quantum superspace approach \cite{z2,z3}.

 The use of quantum superalgebras in physics became popular with the
  introduction of the
deformed harmonic oscillator as a tool
for providing a  realization of the quantum (super) algebra \cite{r25,r26,ku} .  
Particularly, a great  deal of attention  has been paid to  the deformation of 
the  Virasoro algebra \cite{r29,r30,r31,r33,r34,r35}  and  the super 
Virasoro algebra  \cite{r36,r37,z1}.
 Several  investigations of the latter super-algebra are based on the
coordinates 
 and derivatives on super space \cite{z1,r36}. The quantum group strucure on the
 Deformed Virasoro algebra is given in \cite{ma1,ma2} 
 However, the existence of a  quantum super-algebra structure  
of  the deformed super Virasoro algebra is still an open question. The purpose 
of this paper is to give a realization of the deformed super Virasoro 
algebra in terms of bosonic and fermionic algebra operators and define 
a quantum super-algebra  structure on this deformed super-algebra.
In ref. \cite{ma3},  an  algebraic
deformation (associated to an R-matrix of unit square) which allows to a
non trivial Hopf super-algebra structure on the super 
Virasoro and Ramond-Neveu-Shwarz algebras 
 will be
investigated.\\

First, let us  recall that the classical super Virasoro algebra $(sVir)$
 is defined as
an infinite super-algebra generated by the 
 generators $\{L_{n},G_{n},F_{n}; n \in Z \}$ satisfying the defining
relations  \cite{g,rs}
\begin{equation} \label{e1}
\begin{array}{c}
\lbrack L_{n}, L_{m} \rbrack =(n-m) L_{n+m},\\
~~~~\lbrack F_{m}, G_{n} \rbrack =G_{n+m},\\
\lbrack L_{n},F_{m} \rbrack =-m F_{n+m},~~ \\
~~\lbrack F_{m}, F_{n} \rbrack =0,\\
\lbrack L_{m}, G_{n} \rbrack =(m-n) G_{n+m},\\
~~~~\lbrack G_{m}, G_{n} \rbrack =0 ,\\
\end{array}
\end{equation}
 and that the Hopf super-algebra structure on the enveloping super-algebra
of
 this classical super-algebra $U(sVir)$ is trivial. The $Z_2$ 
-grading on this super-algebra is defined by requiring that deg$(L_i) = $ deg$(F_i)
= 0$ and deg$(G_i)= 1$
for $ (i \in Z )$. Further, the bracket $\lbrack, \rbrack$ in (\ref{e1}) stands for a graded one
 $$   \lbrack x, y \rbrack = xy - (-1)^{deg(x) deg(y)} yx. $$
A possible realization of the classical super Virasoro algebra (\ref{e1})
is given as follows
\begin{equation} \label{e2}
\begin{array}{c}
 L_{k}= - (a_{+})^{k+1}a_{-},\\
 G_{n}= (a_{+})^{n+1} f_{+} a_{-} ,\\
 F_{n}= (a_{+})^{n}f_{+} f_{-}
\end{array}
\end{equation}
  where the operators $\{ a_{+}, a_{-}, 1 \} $ generate the classical bosonic algebra
\begin{equation} \label{e3}
\begin{array}{c}
a_{-} a_{+} -  a_{+} a_{-}= 1\\
\lbrack 1 ,a_{\pm} \rbrack =0 .
\end{array}
\end{equation}
and the remaining ones $\{ f_{+}, f_{-}, 1 \}$ generate the classical fermionic 
algebra given by 
\begin{equation} \label{e4}
\begin{array}{c}
f_{-} f_{+} +  f_{+} f_{-}= 1\\
\lbrack 1 ,f_{\pm} \rbrack =0,~~~(f_{\pm})^{2} =0 .
\end{array}
\end{equation}
A q-deformed version of the super Virasoro   algebra  will be defined by 
the generators $\{L_{n},G_{n},K_{n}; n \in Z \}$ and the following
$q$-deformed relations 
\begin{equation} \label{e5}
\begin{array}{c}
 q^{\rm l-k} L_{l} L_{k} - q^{\rm k-l} L_{k} L_{l} = \lbrack l-k \rbrack  L_{k+l}\\
 ~~~~ F_{m} G_{n} - G_{n} F_{m} = G_{n+m},\\
  L_{l} F_{k} - q^{2k} F_{k} L_{l} = - q \lbrack \lbrack k  \rbrack \rbrack F_{k+l},~~~~\\ 
  q^{\rm n-m} F_{m} F_{n} - q^{\rm m-n} F_{n} F_{m}=\lambda \lbrack n-m \rbrack F_{n+m},\\
q^{\rm l-k} L_{l} G_{k} - q^{\rm k-l} G_{k} L_{l} =\lbrack l-k \rbrack G_{k+l},\\
~~~~ G_{m} G_{n} + G_{n} G_{m}  =0 \\
\end{array}
\end{equation}
(where $\lbrack x \rbrack = \frac{q^{x} - q^{-x}}{q - q^{-1}} $, ~~ $\lbrack \lbrack x \rbrack \rbrack = \frac{1 - q^{2x}}{1- q^{2}} $ and $\lambda = q - q^{-1}$).\\ 
The $q$-deformed super-algebra (\ref{e5})
can be realized  in terms of  a q-deformed bosonic algebra generators  $\{ a_{+},
a_{-}, 1 \} $ and a classical 
fermionic algebra generators $\{ f_{+}, f_{-}, 1 \}$ which are such that 
\begin{equation} \label{e6}
\begin{array}{c}
\lbrack a_{-}, a_{+}  \rbrack_{q^{2}} = 1
 ,~~~~\{ f_{-}, f_{+} \} = 1\\
\lbrack a_{\pm} ,f_{\pm} \rbrack =0,~~~~
\lbrack a_{\mp} ,f_{\pm} \rbrack =0,\\ 
(f_{-})^{2} = 0,~~(f_{+})^{2} = 0,\\
\lbrack 1 ,a_{\pm} \rbrack =0, ~~~\lbrack 1 ,f_{\pm} \rbrack =0
\end{array}
\end{equation}
(where $\lbrack a, b \rbrack_{\alpha} = ab - \alpha ba).$\\
 In fact, the following operators 
\begin{equation} \label{e7}
\begin{array}{c}
 L_{k}= - q (a_{+})^{k+1}a_{-},\\
 G_{n}= (a_{+})^{n+1}f_{+} a_{-} ,\\
 F_{n}= (a_{+})^{n}f_{+} f_{-}
\end{array}
\end{equation}
generate the $q$-deformed super Virasoro algebra  (\ref{e5}).\\\\
~~~~~Now we define the  quantum super-algebra structure on the $q$-deformed enveloping 
super-algebra of the $q$-deformed super Virasoro algebra $U_{q}(sVir_{q})$
 as follows:
\begin{equation} \label{e8}
\begin{array}{c}
\Delta (L_{i}) = L_{i} \otimes T_{i} + T_{i} \otimes L_{i},\\
\Delta (G_{i}) = G_{i} \otimes R_{i} + R_{i} \otimes G_{i},\\
\Delta (F_{i}) = F_{i} \otimes K_{i} + K_{i} \otimes F_{i},\\
\epsilon (L_{i}) = 0,\\ 
\epsilon (G_{i}) = 0,\\ 
\epsilon ( F_{i}) = 0,\\
S (L_{i}) = -  S(T_{i}) L_{i} T_{i}^{-1},\\ 
S (G_{i}) = -  S(R_{i}) G_{i} R_{i}^{-1},\\ 
S (F_{i}) = -  S(K_{i}) F_{i} K_{i}^{-1} ,
\end{array}
\end{equation}
where the operators $ \{ K_{i}, T_{i}, R_{i},~~~ i \in Z \} $ are additional
invertible even elements of   $U_{q}(sVir_{q})$ which reduce to the unity
operator when the deformation parameter  $q=1$.\\
From the consistence of the costructures on q-deformed super Virasoro algebra 
generators with the basic axioms of the Hopf super-algebra structure ( co-associativity, 
co-unity and antipode ), we deduce that the costructures on the additional
elements are given by 
\begin{equation} \label{e9}
\begin{array}{c}
\Delta (T_{i}) = T_{i} \otimes T_{i},\\
\Delta (R_{i}) = R_{i} \otimes R_{i},\\
\Delta (K_{i}) = K_{i} \otimes K_{i},\\
\epsilon (T_{i}) =1,\\ 
\epsilon (R_{i}) =1,\\  
\epsilon ( K_{i}) = 1,\\
S (T_{i}) =  T_{i}^{-1},\\ 
S (R_{i}) =  R_{i}^{-1},\\ 
S (K_{i}) =  K_{i}^{-1}.
\end{array}
\end{equation}
Further, the consistence of the Hopf super-algebra structure (\ref{e8}) with
 the
deformed
commutations relations (\ref{e5}), (i.e. using the fact that the coproduct $(\Delta)$
and counit $(\epsilon) $ operations must be  super-algebra homomorphisms, and that 
the coinverse $(S)$ must be  super-algebra anti-homomorphism) we obtain the
following relations between  the $q$-deformed super Virasoro algebra generators and
the additional elements. 
\begin{equation} \label{e10}
\begin{array}{c}
L_{k} T_{l} = q^{2(k+1)l}T_{l} L_{k},\\
L_{l} K_{k} = q^{-1k} K_{k} L_{l},\\ 
G_{k} K_{l} = K_{l} G_{k},\\ 
T_{l} F_{k} = q^{(2+1)k} F_{k}T_{l} ,\\ 
F_{k} R_{l} = R_{l} F_{k},\\ 
G_{k} T_{l} = q^{2(1+k)l} T_{l} G_{k},\\ 
L_{l} R_{k} = q^{2(l+l)k} R_{k} L_{l},\\ 
F_{k} K_{l} = q^{-2(k+1)l} K_{l} F_{k},\\
G_{k} R_{l} =  R_{l} G_{k},\\ 
\end{array}
\end{equation}
 where the additional generators must satisfy the following relations between them
\begin{equation} \label{e11}
\begin{array}{c}
T_{k} T_{l} = T_{l} T_{k} = T_{k+l},\\
T_{k} K_{l} =  K_{l} T_{k} = K_{l+k}\\ 
R_{k} K_{l} = K_{l} R_{k},\\ 
K_{k} K_{l} = K_{l} K_{k} = K_{l+k},\\ 
R_{k} R_{l} = R_{l} R_{k}= R_{k+l},\\ 
T_{l} R_{k} =  R_{k} T_{l}= R_{k+l}.\\
\end{array}
\end{equation}
~~~Considering now the two-parameters $(p,q)$-deformation version of
 the super Virasoro algebra  given by
\begin{equation} \label{e12}
\begin{array}{c}
 q^{\rm l-k} L_{l} L_{k} - q^{\rm k-l} L_{k} L_{l} = \lbrack l-k \rbrack  L_{k+l},\\
 F_{m} G_{n} - G_{n} F_{m} = G_{n+m},\\ 
L_{l} F_{k} - q^{2k} F_{k} L_{l} = - q \lbrack \lbrack k  \rbrack \rbrack F_{k+l},\\
~~~~ q^{\rm n-m} F_{m} F_{n} - q^{\rm m-n} F_{n} F_{m}=\lambda \lbrack n-m \rbrack F_{n+m},\\  
  q^{\rm l-k} L_{l} G_{k} -p^{2l} q^{\rm k-l} G_{k} L_{l} =  \lbrack l-k \rbrack G_{k+l},\\
~~~~ G_{m} G_{n} + G_{n} G_{m}  =0 \\
\end{array}
\end{equation}
which can be realized   as 
\begin{equation} \label{e13}
\begin{array}{c}
 L_{k}= - q (a_{+})^{k+1}a_{-},\\
 G_{n}= p^{-2} (a_{+})^{n+1}f_{+} a_{-} ,\\
 F_{n}= (a_{+})^{n}f_{+} f_{-},
\end{array}
\end{equation}
where the generators $ \{ a_{+}, a_{-}, f_{+}, f_{-},1 \} $ satisfy now  

\begin{equation} \label{e14}
\begin{array}{c}
\lbrack a_{-}, a_{+}  \rbrack_{q^{2}} = 1
 ,~~~~\{ f_{-}, f_{+} \} = 1\\
\lbrack a_{+} ,f_{+} \rbrack_{p^{2}} =0,~~~~
\lbrack a_{-} ,f_{+} \rbrack_{p^{-2}} =0,\\
\lbrack a_{-} ,f_{-} \rbrack_{p^{2}} =0,~~~~
\lbrack a_{+} ,f_{-} \rbrack_{p^{-2}} =0,~~~~\\
(f_{-})^{2} = 0,~~(f_{+})^{2} = 0.\\
\lbrack 1 ,a_{\pm} \rbrack =0,~~~~\lbrack 1 ,f_{\pm} \rbrack =0.
\end{array}
\end{equation}
In this case, the $(p,q)$-quantum super-algebra structure on the $U_{p,q}(sVir_{p,q})$, 
will be given by the same costructures on the $q$-deformed super Virasoro
algebra generators (\ref{e8}) and with the same costructures  on the 
additional even generators. But the deformed
relations between the $(p,q)$-deformed super Virasoro algebra generators and the 
additional elements are 
\begin{equation} \label{e15}
\begin{array}{c}
L_{k} T_{l} = q^{2(k+1)l}T_{l} L_{k},\\
L_{l} K_{k} = q^{(1+l)k} K_{k} L_{l},\\ 
G_{k} K_{l} = K_{l} G_{k},\\ 
F_{k} T_{l} = q^{(k-1)l}T_{l} F_{k},\\ 
F_{k} R_{l} = R_{l} F_{k},\\ 
T_{l} G_{k} = q^{-2(1+k)l} p^{(k+2)l}  G_{k} T_{l} ,\\ 
L_{l} R_{k} = q^{2(l+l)k} p^{-kl}R_{k} L_{l},\\ 
F_{k} K_{l} = q^{-2(k+1)l} K_{l} F_{k},\\
G_{k} R_{l} = R_{l} G_{k},\\ 
\end{array}
\end{equation}
where the additional generators  $ \{ K_{i}, T_{i}, R_{i},~~~ i \in Z \}$
 must satisfy the relations (\ref{e11}) 
as in the one parameter deformation case and reduce to the unity operator
when the deformation parameters p,q reduce to their classical values
 $( p=q=1)$ \\
In  concluding we report that we  established that it is 
possible to define a non trivial Hopf
super-algebra structure on  the
deformed super Virasoro algebra, if we extend  its deformed enveloping 
super-algebra with a set of additional even invertible elements. A consistence realisation
of the additional even invertible elements $ \{ K_{i}, T_{i}, R_{i},~~~ i \in Z \}$
will be given elswere \cite{ma4}
\pagebreak

\end{document}